\documentclass[aps,onecolumn,showpacs,showkeys,nofootinbib]{revtex4}
\usepackage{epsfig}
\usepackage{amsmath}
\usepackage{amsfonts}
\usepackage{amssymb}
\usepackage{graphicx}
\usepackage{colordvi}

\begin{document}

\long\def\symbolfootnote[#1]#2{\begingroup%
\def\thefootnote{\fnsymbol{footnote}}\footnote[#1]{#2}\endgroup}

\title{A general  solution in the Newtonian limit of    $f(R)$ - gravity}
\author{S. Capozziello\footnote{e\,-\,mail address:
capozziello@na.infn.it}$^{\diamond}$,\,A. Stabile\footnote{e -
mail address: stabile@sa.infn.it}$^{\diamond \natural}$, \,A.
Troisi\footnote{e\,-\,mail address:
antrois@gmail.com}$^{\diamond}$}

\affiliation{$^{\diamond}$ Dipartimento di Scienze Fisiche and
INFN, Sez. di Napoli, Universit\`a di Napoli "Federico II", Compl.
Univ. di Monte S. Angelo, Edificio G, Via Cinthia, I-80126 -
Napoli, Italy }

\affiliation{$^{\natural}$ Dipartimento di Ingegneria Meccanica,
Universit\`a di Salerno, via Ponte don Melillo , I- 84084 -
Fisciano (SA), Italy.}

\date{\today}
\begin{abstract}
We show that any analytic $f(R)$-gravity model, in the metric
approach, presents a weak field limit where the standard Newtonian
potential is corrected by a Yukawa-like term. This general result
has never been pointed out but often derived for some particular
theories. This means that only $f(R)=R$ allows to recover the
standard Newton potential while this is not the case for other
relativistic theories of gravity. Some considerations on the
physical consequences of such a general solution are addressed.
\end{abstract}
\maketitle


$f(R)$-gravity can be considered a reliable mechanism to explain
the cosmic acceleration by extending the geometric sector of field
equations
\cite{f(R)-noi1,f(R)-noi2,f(R)-noi3,f(R)-cosmo1,f(R)-cosmo2,f(R)-cosmo3,f(R)-cosmo4,f(R)-cosmo5,palatini1,palatini2,palatini3,palatini4,palatini5,palatini6,palatini7}.
There are several physical and mathematical motivations to enlarge
General Relativity (GR) by these theories. For  comprehensive
reviews, see \cite{GRGreview,OdintsovLadek,farhoudi}.

Besides, dealing with such  extended gravity models at
astrophysical scales, one faces the emergence of corrected
gravitational potentials with respect to the Newton one coming out
from GR. This result is well known since a long time
\cite{stelle,schmidt} but recently it has been pursued to carry
out the possibility to explain the flatness of spiral galaxies
rotation curves and the potential of galaxy clusters without
adding huge amounts of dark matter
\cite{noi-mnras1,noi-mnras2,enzo}.

Other issues as, for example, the observed Pioneer anomaly problem
\cite{anderson1,anderson2} can be framed into the same approach
\cite{bertolami} and then, apart the cosmological dynamics, a
systematic analysis of such theories urges at short scales.

In this letter, we  discuss the Newtonian limit of analytic
$f(R)$-gravity models deriving a  general solution for the
gravitational potential.

The discussion about the short scale behavior of higher order
gravity has been quite vivacious in the last years since  GR shows
its best predictions just at the Solar System level. As matter of
fact, measurements coming from weak field limit tests like the
bending of light, the perihelion shift of planets and the frame
dragging experiments represent inescapable tests for any theory of
gravity. Actually, in our opinion, there are sufficient
theoretical predictions to state that higher order theories of
gravity can be compatible with Newtonian and post-Newtonian
prescriptions \cite{ppn-noi} since the standard Solar System tests
can be evaded by several classes of them \cite{hu}.

Nevertheless, up to now, the discussion on the weak field limit of
$f(R)$\,-\,theories is far to be definitive and there are several
papers claiming for opposite results
\cite{ppn-ok1,ppn-ok2,ppn-ok3,ppn-no1,ppn-no2,ppn-no3,ppn-no4,ppn-no5},
or stating that no progress has been reached in the last  forty
years due to several misconceptions in the various theories of
gravity \cite{faraonithomas1,faraonithomas2}.

In particular, people approached the weak limit issue following
different schemes and developing different parameterizations
which, in some cases, turn out to be not necessarily correct.

The present analysis is based on the metric approach, developed in
the Jordan frame, assuming that the observations are performed in
it, without resorting to any conformal transformation (see also
\cite{noi-prd}). This point of view is adopted in order to avoid
dangerous variable changes which could compromise the correct
physical interpretation of the results. As we will see below, a
general gravitational potential, with a Yukawa correction, can be
achieved in the Newtonian limit of any analytic $f(R)$-gravity
model. From a phenomenological point of view, this correction
allows to consider as viable this kind of models even at small
distances, provided that the Yukawa correction turns out to be
insignificant in this approximation as in the so called "chameleon
mechanism" \cite{chameleon}. From the point of view of the
post-Newtonian corrections, such a solution implies  a
redefinition of the coupling constants in order to fulfill  the
experimental prescriptions.

As matter of fact, if one  evaluates what  the corrections are to
the Newtonian potential coming from a modified gravity model in a
post-Newtonian regime, it is necessary to take into account
corrections both at the second order and at the fourth order in
the perturbation expansion of the metric. Furthermore, this
analysis turns out to be coherent only if a small $r$ regime is
adopted. In this limit, in fact, the potential can be retained
reasonably Newtonian-like in relation with experimental results.

Let us start from  a general fourth order gravity action \,:

\begin{eqnarray}\label{actfR}
\mathcal{A}\, = \,\int
d^4x\sqrt{-g}\biggl[f(R)+\mathcal{X}\mathcal{L}_m\biggr]\,,
\end{eqnarray}
where $f(R)$ is an  analytic function of Ricci scalar, $g$ is the
determinant of the metric $g_{\mu\nu}$, $\mathcal{X}=\frac{16\pi
G}{c^4}$ is the coupling constant and $\mathcal{L}_m$ describes
the standard fluid-matter Lagrangian. It is well known that such
an action is the straightforward generalization of the
Hilbert-Einstein action of GR obtained for $f(R)=R$. Since we are
considering the metric approach, field equations are obtained by
varying (\ref{actfR}) with respect to the metric\,:

\begin{eqnarray}\label{fe}
f'R_{\mu\nu}-\frac{1}{2}fg_{\mu\nu}-f'_{;\mu\nu}+g_{\mu\nu}\Box
f'=\frac{\mathcal{X}}{2}T_{\mu\nu}\,,
\end{eqnarray}
and the trace is
\begin{eqnarray}\label{fetr}
3\Box f'+f'R-2f=\frac{\mathcal{X}}{2}T\,.
\end{eqnarray}
${\displaystyle
T_{\mu\nu}=\frac{-2}{\sqrt{-g}}\frac{\delta(\sqrt{-g}\mathcal{L}_m)}{\delta
g^{\mu\nu}}}$ is the energy momentum tensor of matter ($T$ is its
trace), ${\displaystyle f'=\frac{df(R)}{dR}}$ and
$\Box={{}_{;\sigma}}^{;\sigma}$ is the d'Alembert operator.

Actually, as discussed in \cite{noi-prd}, we deal with the
Newtonian and the post-Newtonian limit of $f(R)$ - gravity
adopting the  spherical symmetry. The solution of field equations
can be obtained considering the metric\,:

\begin{eqnarray}\label{me}
ds^2\,=\,g_{\sigma\tau}dx^\sigma
dx^\tau=g_{00}(x^0,r)d{x^0}^2+g_{rr}(x^0,r)dr^2-r^2d\Omega
\end{eqnarray}
where $x^0\,=\,ct$ and $d\Omega$ is the angular element.

In order to develop the Newtonian limit, let us consider the
perturbed metric with respect to a Minkowskian background
$g_{\mu\nu}\,=\,\eta_{\mu\nu}+h_{\mu\nu}$. The metric entries can
be developed as:
\begin{eqnarray}\label{definexpans}
\left\{\begin{array}{ll} g_{tt}(t,
r)\simeq1+g^{(2)}_{tt}(t,r)+g^{(4)}_{tt}(t,r)
\\\\
g_{rr}(t,r)\simeq-1+g^{(2)}_{rr}(t,r)\\\\
g_{\theta\theta}(t,r)=-r^2\\\\
g_{\phi\phi}(t,r)=-r^2\sin^2\theta
\end{array}\right.\,,
\end{eqnarray}
where we are assuming $c\,=\,1$\,,\ $x^0=ct\rightarrow t$\, and
applying the  formalism developed in \cite{noi-prd}.

Since we want to obtain the most general result, we does not
provide any specific form for the $f(R)$-Lagrangian. We assume,
however,  analytic Taylor expandable $f(R)$ functions with respect
to a certain value $R\,=\,R_0$\,:

\begin{eqnarray}\label{sertay}
f(R)=\sum_{n}\frac{f^n(R_0)}{n!}(R-R_0)^n\simeq
f_0+f_1R+f_2R^2+f_3R^3+...\,.
\end{eqnarray}

In order to obtain the post-Newtonian approximation, one has to
insert  expansions (\ref{definexpans}) and (\ref{sertay}) into
field equations (\ref{fe}) - (\ref{fetr}) and expand the system up
to the orders $O(0)$, $O(2)$ e $O(4)$. This approach provides
general results and specific (analytic) theories are selected by
the coefficients $f_i$ in Eq.(\ref{sertay}). It is worth noting
that, at the order $O(0)$, the field equations give the condition
$f_0 =0$ and then the solutions at further orders do not depend on
this parameter as we will show below. If we now consider the
$O(2)$ - order approximation, the equations system (\ref{fe}) in
vacuum, results to be

\begin{eqnarray}\label{eq2}
\left\{\begin{array}{ll}
f_1rR^{(2)}-2f_1g^{(2)}_{tt,r}+8f_2R^{(2)}_{,r}-f_1rg^{(2)}_{tt,rr}+4f_2rR^{(2)}=0
\\\\
f_1rR^{(2)}-2f_1g^{(2)}_{rr,r}+8f_2R^{(2)}_{,r}-f_1rg^{(2)}_{tt,rr}=0
\\\\
2f_1g^{(2)}_{rr}-r[f_1rR^{(2)}-f_1g^{(2)}_{tt,r}-f_1g^{(2)}_{rr,r}+4f_2R^{(2)}_{,r}+4f_2rR^{(2)}_{,rr}]=0
\\\\
f_1rR^{(2)}+6f_2[2R^{(2)}_{,r}+rR^{(2)}_{,rr}]=0
\\\\
2g^{(2)}_{rr}+r[2g^{(2)}_{tt,r}-rR^{(2)}+2g^{(2)}_{rr,r}+rg^{(2)}_{tt,rr}]=0
\end{array} \right.\end{eqnarray}
It is evident that the trace equation (the fourth in the system
(\ref{eq2})), provides a differential equation with respect to the
Ricci scalar which allows to solve  exactly the system (\ref{eq2})
at $O(2)$ - order. Finally, one gets the general solution\,:

\begin{eqnarray}\label{sol}
\left\{\begin{array}{ll}
g^{(2)}_{tt}=\delta_0-\frac{Y}{f_1r}-\frac{\delta_1(t)e^{-r\sqrt{-\xi}}}{3\xi
r}+\frac{\delta_2(t)e^{r\sqrt{-\xi}}}{6({-\xi)}^{3/2}r}
\\\\
g^{(2)}_{rr}=-\frac{Y}{f_1r}+\frac{\delta_1(t)[r\sqrt{-\xi}+1]e^{-r\sqrt{-\xi}}}{3\xi
r}-\frac{\delta_2(t)[\xi r+\sqrt{-\xi}]e^{r\sqrt{-\xi}}}{6\xi^2r}
\\\\
R^{(2)}=\frac{\delta_1(t)e^{-r\sqrt{-\xi}}}{r}-\frac{\delta_2(t)\sqrt{-\xi}e^{r\sqrt{-\xi}}}{2\xi
r}\end{array} \right.
\end{eqnarray}
where $\xi\doteq\displaystyle\frac{f_1}{6f_2}$, $f_1$ and $f_2$
are the expansion coefficients obtained by Taylor developing the
analytic $f(R)$ Lagrangian and $Y$ is an arbitrary integration
constant. When we consider the limit $f\rightarrow R$, in the case
of a point-like source of mass $M$, we recover the standard
Schwarzschild solution with $Y=2GM$. Let us notice that the
integration constant $\delta_0$ is  dimensionless, while the two
arbitrary functions of time $\delta_1(t)$ and $\delta_2(t)$ have
respectively the dimensions of $lenght^{-1}$ and $lenght^{-2}$;
$\xi$ has the dimension $lenght^{-2}$.  The functions of time
$\delta_i(t)$ ($i=1,2$) are completely arbitrary since the
differential equation system (\ref{eq2}) contains only spatial
derivatives and can be settled to  constant values. Besides, the
integration constant $\delta_0$ can be set to zero since it
represents an unessential additive quantity for the potential.

We can now write  the general solution of the problem considering
the previous expression (\ref{sol}). In order to match at infinity
the Minkowskian prescription for the metric, we can discard the
Yukawa growing mode  in (\ref{sol}) and then we have\,:

\begin{eqnarray}\label{mesol}
\left\{\begin{array}{ll}ds^2\,=\,\biggl[1-\frac{2GM}{f_1r}-\frac{\delta_1(t)e^{-r\sqrt{-\xi}}}{3\xi
r}\biggr]dt^2-\biggl[1+\frac{2GM}{f_1r}-\frac{\delta_1(t)(r\sqrt{-\xi}+1)e^{-r\sqrt{-\xi}}}{3\xi
r}\biggr]dr^2-r^2d\Omega\\\\R\,=\,\frac{\delta_1(t)e^{-r\sqrt{-\xi}}}{r}\end{array}\right.
\end{eqnarray}
At this point, one can provide the solution in term of the
gravitational potential.  The first of (\ref{sol}) gives the
second order solution in term of the metric expansion (see the
definition (\ref{definexpans})). This term coincides with the
gravitational potential at the Newton order. In particular, since
 $g_{tt}\,=\,1+2\Phi_{grav}\,=\,1+g_{tt}^{(2)}$, the general
gravitational potential of $f(R)$-gravity, analytic in the Ricci
scalar $R$, is

\begin{eqnarray}\label{gravpot}
\Phi_{grav}\,=\,-\left(\frac{GM}{f_1r}+\frac{\delta_1(t)e^{-r\sqrt{-\xi}}}{6\xi
r}\right)\,.
\end{eqnarray}
This general result means that the standard Newton potential is
achieved only in the particular case $f(R)=R$ while it is not so
for generic analytic $f(R)$ models. Eq.(\ref{gravpot}) deserves
some comments. The parameters $f_{1,2}$ and the function
$\delta_1$ represent the deviations with respect the standard
Newton potential. To test such theories of gravity inside the
Solar System, we need to compare such quantities with respect to
the current experiments, or, in other words, Solar System
constraints  should be evaded fixing such parameters. On the other
hand, such parameters could acquire  non-trivial values (e.g.
$f_1\neq 1,\,\delta_1(t)\neq 0,\,\xi\neq 1$) at   scales different
from the Solar System ones.

In conclusions, we have derived the Newtonian limit for analytic
$f(R)$-gravity models. The approach which we have developed can be
adopted also to  calculate correctly the post-Newtonian parameters
of such theories without any redefinition of the degrees of
freedom by some scalar field. This last step  could be misleading
in the weak field limit. In the approach presented here, we do not
need any change from the Jordan to the Einstein frame
\cite{ppn-noi-bis,sotiriou2}.

Apart the possible shortcomings related to  non-correct changes of
variables, any $f(R)$ theory can be rewritten  as a scalar-tensor
one or  an ideal fluid, as shown in \cite{nodi,nodicap1,nodicap2}.
In those papers, it has been demonstrated that such different
representations give rise to physically non-equivalent theories
and then also the Newtonian and post-Newtonian approximations have
to be handled very carefully since the results could not be
equivalent. In fact, the further geometric degrees of freedom of
$f(R)$ gravity (with respect to GR), namely the scalar field
and/or the further ideal fluid, have weak field behaviors strictly
depending on the adopted gauge which could not be equivalent or
difficult to compare. In order to circumvent these possible
sources of shortcomings, one should states the frame (Jordan or
Einstein) at the very beginning and then remain in such a frame
along all the calculations up to the final results. Adopting this
procedure, arbitrary limits and non-compatible results should be
avoided.

As we have seen, the solution relative to the $g_{tt}$ metric
component gives the gravitational potential  corrected by
Yukawa-like terms, combined with the Newtonian potential. In
relation to the sign of the  coefficients entering  $g_{tt}$, one
can obtain real or complex solutions with different physical
meanings. This degeneracy could be removed once standard matter is
introduced into dynamics \cite{chameleon}. On the other hand, the
growing Yukawa mode can be neglected as soon as the correct
asymptotic Minkowskian limit is required. Such considerations
acquire an important meaning as soon as $f(R)$ gravity is
considered from the viewpoints of stability and Cauchy problem. It
is possible to show that the correct Newtonian limit of the theory
is strictly related to the stability of the ground state solution
\cite{faraoni1,faraoni2,leszek,straumann} and this fact determines
the range of viable parameters in the potential (\ref{gravpot}).

\end{document}